\documentclass[sigconf,authordraft, review=false]{acmart}

\AtBeginDocument{%
  \providecommand\BibTeX{{%
    \normalfont B\kern-0.5em{\scshape i\kern-0.25em b}\kern-0.8em\TeX}}}

\acmConference[arXiv]{e-Print Archive}{2019}{USA}
\begin{document}

% \title{The Name of the Title is Hope}
\title{CupQ: A New Clinical Literature Search Engine}

\author{Jesse Wang}
\orcid{0000-0001-8269-1930}
\affiliation{%
    \institution{Department of Translational Biomedical Science\\
    School of Medicine and Dentistry\\
    University of Rochester}
    \city{Rochester}
    \state{NY}
    \country{USA}
}
\email{jesse_wang@urmc.rochester.edu}

\author{Henry Kautz}
\orcid{0000-0001-5219-2970}
\affiliation{%
    \institution{Department of Computer Science\\
    Hajim School of Engineering and Applied Sciences\\
    University of Rochester}
    \city{Rochester}
    \state{NY}
    \country{USA}
}
\email{kautz@cs.rochester.edu}

%% document.
\begin{abstract}
A new clinical literature search engine, called CupQ, is presented. It aims to help clinicians stay updated with medical knowledge. Although PubMed is currently one of the most widely used digital libraries for biomedical information, it frequently does not return clinically relevant results. CupQ utilizes a ranking algorithm that filters non-medical journals, compares semantic similarity between queries, and incorporates journal impact factor and publication date. It organizes search results into useful categories for medical practitioners: reviews, guidelines, and studies. Qualitative comparisons suggest that CupQ may return more clinically relevant information than PubMed. CupQ is available at https://cupq.io/.
\end{abstract}




\maketitle

\section{Introduction}
The task of staying updated with advances in medicine remains a challenging aspect of clinical practice. An average of about two biomedical documents is added to the literature every minute \cite{fiorini2018best}. The widely used PubMed digital library often does not deliver clinically relevant results within a reasonable time frame \cite{agoritsas2014increasing,ho2016development,davies2011physicians}. Other resources, such as UpToDate, NEJM Journal Watch, and ACP Journal Club, rely on the expensive and time-consuming process of using human curators to manually comb the literature for clinical information \cite{uptodate,nejm,acp}. The current utilities for medical information retrieval may be inadequate for continuing medical education and consequently may be hindering efforts to improve patient care.

PubMed is a biomedical digital library built and maintained by the United States National Center for Biotechnology Information \cite{pubmed}. It often requires users to select filters, identify MeSH terms, and generate boolean entries to distill relevant clinical results \cite{russell2017expert,lindsey2013pubmed}. The complexity of PubMed may contribute to low search satisfaction among healthcare professionals \cite{agoritsas2014increasing,ho2016development,davies2011physicians}. Moreover, the newly released Best Match relevance algorithm does not incorporate important metrics such as journal rank and semantic similarity \cite{fiorini2018best}. These ranking signals also appear to be missing in the related search tool, PubMed Clinical Queries \cite{pubmedclinicalqueries}. To better fulfill the information needs of medical practice, PubMed may require further improvements.

This application note discusses the development of a new medical literature search engine called CupQ. The system uses Word2Vec to generate word embeddings for comparing semantic similarity between queries and documents \cite{mikolov2013distributed,white2015well}. It also considers journal impact factor (JIF) and publication date \cite{jcrclarivate}. Results are organized by reviews, guidelines, and clinical studies. Documents written in English and published in journals listed in the medicine subject area of ScimagoJR are returned \cite{scimagojr}. Example search results suggest that CupQ may be more effective than PubMed for returning relevant clinical information. This publication aims to encourage utilization of CupQ for staying updated with medical literature.

\section{Related Work}
Lu provides a survey of web tools for searching biomedical literature, including Quertle, MEDIE, and Semantic MEDLINE \cite{lu2011pubmed}.

Quertle is a semantic search engine utilizing over 250 million subject-verb-object (SVO) associations to provide relevant publications \cite{rindflesch2011semantic}. It also features "Power Terms" that allow users to search topics. Example terms, denoted by a dollar sign (\$) prefix, include "\$Amino Acids," "\$Biomarkers," and "\$Chemicals." In addition, Quertle differentiates capitalizations, such as the "WHO" abbreviation for the World Health Organization and the "who" pronoun. Search results are presented in two tabs. One tab lists results derived from its semantic-based algorithm. Another tab lists results obtained from a standard PubMed search. Quertle was developed and is currently maintained by a for-profit private enterprise. The exact details of its search process are consequently unavailable to the public.

MEDIE also aims to incorporate grammatical meaning into its search algorithm \cite{ohta2006intelligent}. It returns documents that match the user's desired SVO relations. For example, the query "what does p53 activate" would produce results that contain sentences matching "activate" and "p53" as the verb and object, respectively. Queries in MEDIE are first annotated with part-of-speech tags through the Enju head-driven phrase structure grammar parser. Genes and diseases are also annotated through a dictionary comparison approach. After annotation, results returned from a standard keyword search are filtered based on the predicate structure of their sentences. Users are shown the results with sentences matching the specified semantic relations.

Semantic MEDLINE, similar to Quertle and MEDIE, utilizes linguistic information \cite{rindflesch2011semantic}. In particular, it extracts normalized representations of semantic relations. For example, the phrase "Genes AFFECTS Circadian Rhythms" was parsed from the title "Clock genes are the genes that control circadian rhythms in physiology and behavior." The extraction process was developed using the SemRep natural language processing platform, which depends on the National Library of Medicine's Unified Medical Language System. Semantic categories and relationships are derived from this collection. The process was conducted on about 25 million MEDLINE abstracts and produced more than 26 million semantic relations.

\section{Methods and Materials}

\subsection{Server Architecture}
An instance of CupQ uses two networked servers. A dedicated storage server is used to maintain persistent information, including a MySQL database, a MongoDB database, and other files. The storage server also performs operations relating to data download and extraction. Another server containing high memory capacity is used for tokenization, embedding, indexing, searching, and website hosting. The storage server contains an Intel Core i7-4790K 4.0 GHz processor, Ballistix Sport 32 GB DDR3 RAM, and a Samsung 850 Evo 1 TB SSD. The memory server is a Dell R710 with dual Intel Xeon X5687 3.6 GHz processors, 288 GB PC3-10600R RAM, and a Samsung 850 Evo 256 GB SSD.

\subsection{Data Download and Extraction}
MEDLINE/PubMed data is downloaded via FTP as a directory of compressed XML files \cite{pubmeddownload}. MD5 checksums are compared to ensure file integrity. Specific XML elements related to title, abstract, journal, authors, and publication date are parsed and inserted into a MongoDB collection. The most recent journal information from ScimagoJR and Journal Citation Reports is also downloaded. Documents published in journals listed in the medicine subject area of ScimagoJR are labeled. Each document in this subset is assigned the JIF of its publishing journal. Subsequent operations are performed only on this document subset.

\subsection{Tokenization and Embedding}
Tokens are extracted from titles and abstracts by splitting text on space and hyphen characters. The LuiNorm API is used for token normalization \cite{lexicaltools}. Stopwords, except for those fully capitalized, are removed. Then, the Genism library is used to run Word2Vec, generating a vector representation for each token \cite{gensim}. Vectors of 100 elements are produced using skip-gram and a window size of 100 without sentence boundaries for 10 epochs. Embeddings for document titles are computed as the sum of each token embedding multiplied by the log ratio of the corpus size to the number of documents containing the token. 

\subsection{Inverted Indexing}
A Java hash map with keys as integers and values as integer array lists is instantiated. Keys represent numeric token identifiers (TIDs). Values represent document PubMed identifiers (PMIDs). For each document, a hash set of title and abstract TIDs is created. The document PMID is added to the array list for each TID. Key-value pairs are stored in a MySQL table comprised of two integer columns, the first for TIDs and the second for PMIDs, with the primary key set over both columns. New MEDLINE/PubMed documents are automatically downloaded, processed, and indexed on a weekly basis.

\subsection{Document Retrieval}
A Java search server tokenizes the search string and computes a weighted sum vector representation. The token contained in the fewest documents is passed to the inverted index, which returns a list of PMIDs. Only results written in English and containing all search tokens are retained. Errata, retracted documents, and documents published before the year 1990 are removed. Document information, including publication date, publication type, title embedding, and TIDs are stored in an object array. Results are organized by publication type into array lists for reviews, guidelines, and studies.

After assigning documents into publication categories, a relevance score is computed for each document. A different relevance calculation is used for each publication category. Document lists are then sorted by relevance in descending order. The top 500 results are retained and cached into a MySQL table. A sublist containing results to be displayed for the user's requested page number is obtained. Display information including title, abstract, author abbreviations, journal ISO abbreviation, and publication year is retrieved from disk. Search results are returned to the web server as a JSON payload for HTML rendering.

The document relevance score is the sum of several min-max normalized subscores multiplied by empirically configured boosting factors. A semantic score is computed as the cosine similarity between the query vector and the title vector. A title count score is set to one if a title contains all search tokens and zero otherwise. A date score is computed as an estimated number of days. A journal score is set to the JIF. If a document is published over twenty years ago, the relevance score is fractioned by a tenth. If any of the subscores are zero, then the relevance score is zero. Different sets of boosting factors are used for each category (Table \ref{tab:boostingfactors}).

\begin{table}[H]
  \caption{Publication Category Boosting Factors}
  \label{tab:boostingfactors}
  \begin{tabular}{lcccc}
    \toprule
    Category & Title Cosine & Title Count & Date & Journal\\
    \midrule
    Reviews & 4 & 3 & 1 & 2\\
    Guidelines & 6 & 8 & 1 & 4\\
    Studies & 3 & 5 & 1 & 1\\
  \bottomrule
\end{tabular}
\end{table}

\section{Results}

\subsection{User Interface}
\begin{figure}[h]
\centering
\includegraphics[width=\linewidth]{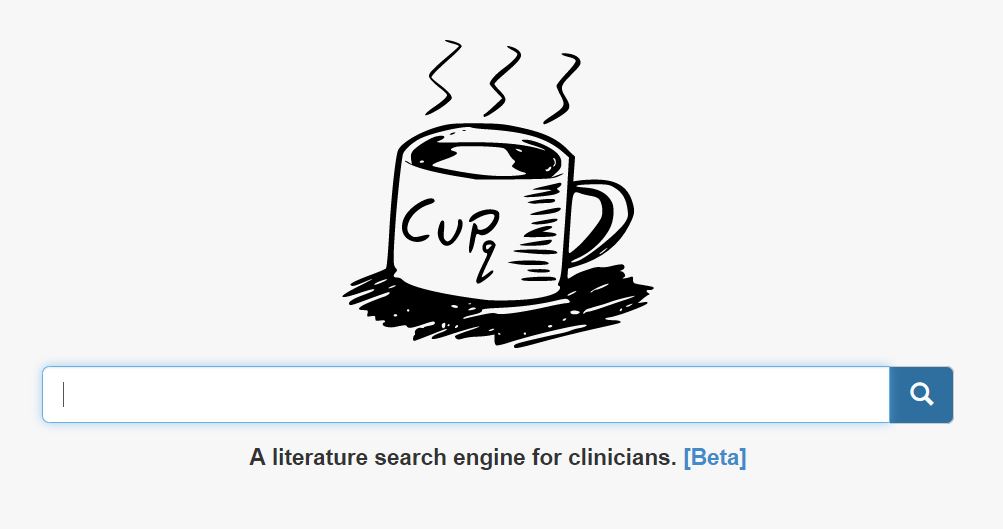}
\caption{CupQ home page.}
\label{fig:1}  
\end{figure}

\begin{figure}[h]
\centering
\includegraphics[width=\linewidth]{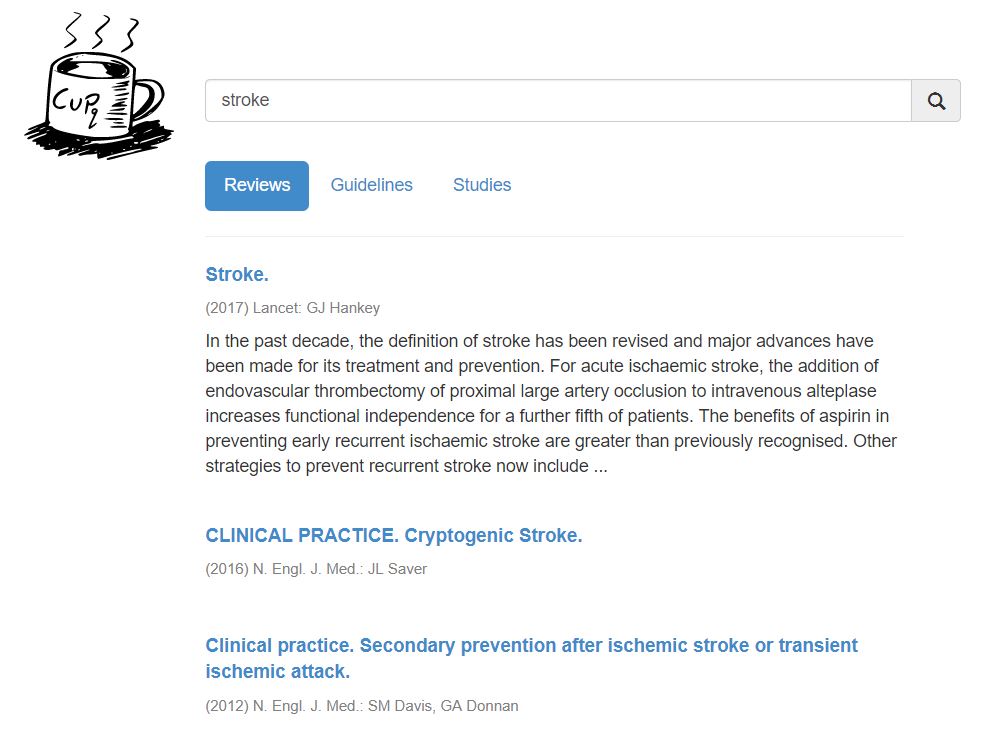}
\caption{CupQ results page showing the top three results for the query "stroke."}
\label{fig:2}
\end{figure}

CupQ provides a simple user interface that includes a search bar for entering queries and a tab bar for selecting publication categories (Figures \ref{fig:1}--\ref{fig:2}).

\subsection{Search Results}
The top ten results for several queries and filters were compared between CupQ and PubMed. Note that the PubMed sidebar does not allow for selection of document types to be excluded. For example, it does not allow inclusion of documents that are reviews but not systematic reviews. Defining document types using advanced search strings in PubMed returns different results than selecting document types in the sidebar, perhaps because the Best Match ranking algorithm weighs document types in the search string whereas the sidebar selection behaves as a simple binary filter. Search result comparisons were made with the PubMed sidebar because its interface is more similar to the CupQ tab bar. Searches were performed on January 28, 2019.

\subsubsection{Myocardial Infarction (Reviews)}
For the query "myocardial infarction," CupQ returned review results from high impact factor journals, including \textit{New England Journal of Medicine} (JIF = 79.26), \textit{Lancet} (JIF = 53.254), and \textit{BMJ} (JIF = 23.562) (Table \ref{tab:cupqmi}). The titles of the first two results, "Acute myocardial infarction," were highly relevant to the query with a cosine similarity of 0.986. Both documents were published in 2017 issues of \textit{New England Journal of Medicine} and \textit{Lancet}. Other results referenced common concepts related to myocardial infarction, including coronary reperfusion strategies, percutaneous coronary intervention, electrocardiogram, and ST-segment elevation. Incorporation of JIF and Word2Vec query-title cosine similarity may explain the effective prioritization of high impact factor journals and titles containing semantically related concepts to myocardial infarction.

PubMed contrastingly returned no results from \textit{New England Journal of Medicine}, \textit{Lancet}, or \textit{BMJ} (Table \ref{tab:pubmedmi}). The title of the first result was less relevant to the query with a cosine similarity of 0.832. Moreover, the first result was published in an unranked journal by JIF. Although there was a document with the highly relevant title "Acute myocardial infarction," it was published in a 2013 issue of \textit{Disease-A-Month}, a relatively low impact factor journal (JIF = 0.891). PubMed did not return the newer documents with the same title from \textit{New England Journal of Medicine} and \textit{Lancet}. In addition, only one PubMed title referenced an aforementioned common topic related to myocardial infarction, percutaneous coronary intervention. PubMed and CupQ shared no common results. These observations suggest that PubMed may not effectively incorporate JIF in the context of query-title semantic similarity.

\subsubsection{Depression (Guidelines)}
There were more similarities between CupQ and PubMed for the query "depression" when searching for guidelines (Tables \ref{tab:cupqdepression}--\ref{tab:pubmeddepression}). The same document published in a 2016 issue of \textit{JAMA} (JIF = 47.661) appeared as the top result for both search engines. However, PubMed lacked the result "Screening for Depression in Children and Adolescents: U.S. Preventive Services Task Force Recommendation Statement," published in a 2016 issue of \textit{Annals of Internal Medicine} (JIF = 19.384). This was unusual behavior because PubMed was able to return a document with the same title and year, albeit from a lower impact factor journal, \textit{Pediatrics} (JIF = 5.515). Unlike PubMed, CupQ may return relevant results by estimating importance via JIF.

All result titles in CupQ contained the query "depression." A problem with PubMed was that the title of the fourth result did not contain the query. Although this result was published within the last two years in a high impact factor journal, \textit{CA: A Cancer Journal for Clinicians} (JIF = 244.585), it did not specifically focus on depression. This document encompassed strategies for addressing multiple conditions in patients with breast cancer, including chemotherapy-induced nausea, vomiting, and peripheral neuropathy. Although this document may be more appropriate as a top ten result for the query "depression breast cancer," it addresses too many topics other than depression to be a top ten result for the query "depression."

\subsubsection{Stroke (Studies)}
When searching for studies about stroke, all result titles from CupQ and PubMed contained the query (Tables \ref{tab:cupqstroke}--\ref{tab:pubmedstroke}). CupQ only returned results from \textit{New England Journal of Medicine} whereas PubMed returned no results from this journal. The first result returned by PubMed was published in \textit{Clinical Neurology and Neurosurgery} (JIF = 1.736). The highest impact factor journal returned by PubMed was \textit{Lancet Neurology} (JIF = 27.144). The first result from CupQ was published in 2018 whereas the first result from PubMed was published 2017. Moreover, CupQ results were published from 2017 to 2018 whereas PubMed results were published from 2007 to 2018. CupQ may prioritize recent, high impact factor results whose titles contain the query.

\section{Discussion and Conclusion}
Search engine performance can be assessed through a variety of approaches. Precision and recall can be measured, assuming a binary relevance model and an existing standard for relevance \cite{hawking2001measuring}. User task studies may demonstrate performance with respect to specific search objectives but may require statistical adjustment for prior user experience with comparative search tools \cite{taksa2008task}. Click-through rates may provide another indication of performance given a high volume of web traffic \cite{fiorini2018best}. This paper qualitatively compared results between CupQ and PubMed for specific queries and publication categories. Because CupQ was recently launched in January 2019, future work will include analyses of click-through rates when there is significant traffic.

The CupQ ranking algorithm prioritizes title relevance, JIF, and publication date. It assumes that users place the most emphasis on title content when determining the relevance of a result. It also assumes that users weigh the reliability and importance of information, represented by JIF, either greater than or equal to the recency of information. Although JIF is not necessarily representative of individual articles in a journal, it may serve as a useful approximation for physicians who may have limited time to search for information \cite{garfield2006history,seglen1997impact,saha2003impact}. In addition, CupQ only returns information published in journals that are listed in the medicine subject area of ScimagoJR. This unique implementation of title relevance, JIF, publication date, and journal category may enable CupQ to return relevant clinical information.

%% identification of the section in the document metadata, and the
\begin{acks}
Jesse Wang is an MD and PhD candidate in the Medical Scientist Training Program funded by the National Institute of Health under grant T32 GM07356. The content is solely the responsibility of the author and does not necessarily represent the official views of the National Institute of General Medicine Science or the National Institute of Health. We thank Jie Wang at the University of Massachusetts Lowell and Daniel Schwartz at the University of Connecticut for their comments that greatly improved this manuscript.
\end{acks}

\bibliographystyle{ACM-Reference-Format}
\bibliography{my-citations}

\appendix

\section*{Appendix}
The appendix consists of Tables A1--A6.

\setcounter{table}{0}
\renewcommand{\thetable}{A\arabic{table}}

\begin{table*}
  \caption{Reviews returned by CupQ for the query "myocardial infarction."}
  \label{tab:cupqmi}
  \begin{tabular}{|l|p{10cm}|p{5cm}|l|}
    \hline
    \textbf{No} & \textbf{Title} & \textbf{Journal} & \textbf{Year}\\
    \hline
    1 & Acute Myocardial Infarction. & New England Journal of Medicine & 2017\\
    \hline
    2 & Acute myocardial infarction. & Lancet & 2017\\
    \hline
    3 & Myocardial infarction due to percutaneous coronary intervention. & New England Journal of Medicine & 2011\\
    \hline
    4 & Primary PCI for myocardial infarction with ST-segment elevation. & New England Journal of Medicine & 2007\\
    \hline
    5 & Acute myocardial infarction. & Lancet & 2008\\
    \hline
    6 & Use of the electrocardiogram in acute myocardial infarction. & New England Journal of Medicine & 2003\\
    \hline
    7 & Future treatment strategies in ST-segment elevation myocardial infarction. & Lancet & 2013\\
    \hline
    8 & Reperfusion strategies in acute myocardial infarction and multivessel disease. & Nature Reviews Cardiology & 2017\\
    \hline
    9 & Coronary microvascular obstruction in acute myocardial infarction. & European Heart Journal & 2016\\
    \hline
    10 & Management of patients after primary percutaneous coronary intervention for myocardial infarction. & BMJ & 2017\\
    \hline
  \end{tabular}
\end{table*}

\begin{table*}
  \caption{Reviews returned by PubMed for the query "myocardial infarction."}
  \label{tab:pubmedmi}
  \begin{tabular}{|l|p{10cm}|p{5cm}|l|}
    \hline
    \textbf{No} & \textbf{Title} & \textbf{Journal} & \textbf{Year}\\
    \hline
    1 & Myocardial infarction with non obstructive coronary arteries (MINOCA): a whole new ball game. & Expert Review of Cardiovascular Therapy & 2017\\
    \hline
    2 & Type 2 myocardial infarction due to supply-demand mismatch. & Trends in Cardiovascular Medicine & 2017\\
    \hline
    3 & Assessment and classification of patients with myocardial injury and infarction in clinical practice. & Heart & 2017\\
    \hline
    4 & Multivessel versus culprit lesion only percutaneous coronary intervention in cardiogenic shock complicating acute myocardial infarction: A systematic review and meta-analysis. & European Heart Journal: Acute Cardiovascular Care & 2018\\
    \hline
    5 & Exosomes and cardiac repair after myocardial infarction. & Circulation Research & 2014\\
    \hline
    6 & Acute myocardial infarction. & Disease-A-Month & 2013\\
    \hline
    7 & Perioperative myocardial infarction/injury after noncardiac surgery. & Swiss Medical Weekly & 2015\\
    \hline
    8 & MicroRNAs in myocardial infarction. & Nature Reviews Cardiology & 2015\\
    \hline
    9 & Galectin-3 and post-myocardial infarction cardiac remodeling. & European Journal of Pharmacology & 2015\\
    \hline
    10 & Type 2 myocardial infarction: the chimaera of cardiology? & Heart & 2015\\
    \hline
  \end{tabular}
\end{table*}

\begin{table*}
  \caption{Guidelines returned by CupQ for the query "depression."}
  \label{tab:cupqdepression}
  \begin{tabular}{|l|p{10cm}|p{5cm}|l|}
    \hline
    \textbf{No} & \textbf{Title} & \textbf{Journal} & \textbf{Year}\\
    \hline
    1 & Screening for Depression in Adults: US Preventive Services Task Force Recommendation Statement. & JAMA & 2016\\
    \hline
    2 & Confronting depression and suicide in physicians: a consensus statement. & JAMA & 2003\\
    \hline
    3 & Screening for Depression in Children and Adolescents: U.S. Preventive Services Task Force Recommendation Statement. & Annals of Internal Medicine & 2016\\
    \hline
    4 & Screening for depression in adults: U.S. preventive services task force recommendation statement. & Annals of Internal Medicine & 2009\\
    \hline
    5 & Screening for Depression in Children and Adolescents: US Preventive Services Task Force Recommendation Statement. & Pediatrics & 2016\\
    \hline
    6 & European Psychiatric Association Guidance on psychotherapy in chronic depression across Europe. & European Psychiatry & 2016\\
    \hline
    7 & Management of Depression in Patients With Cancer: A Clinical Practice Guideline. & Journal of Oncology Practice & 2016\\
    \hline
    8 & Screening for Depression in Adults: Recommendation Statement. & American Family Physician & 2016\\
    \hline
    9 & Evidence-based interventions to improve the palliative care of pain, dyspnea, and depression at the end of life: a clinical practice guideline from the American College of Physicians. & Annals of Internal Medicine & 2008\\
    \hline
    10 & Clinical pathway for the screening, assessment and management of anxiety and depression in adult cancer patients: Australian guidelines. & Psycho-oncology & 2015\\
    \hline
  \end{tabular}
\end{table*}

\begin{table*}
  \caption{Guidelines returned by PubMed for the query "depression."}
  \label{tab:pubmeddepression}
  \begin{tabular}{|l|p{10cm}|p{5cm}|l|}
    \hline
    \textbf{No} & \textbf{Title} & \textbf{Journal} & \textbf{Year}\\
    \hline
    1 & Screening for Depression in Adults: US Preventive Services Task Force Recommendation Statement. & JAMA & 2016\\
    \hline
    2 & Consensus Recommendations for the Clinical Application of Repetitive Transcranial Magnetic Stimulation (rTMS) in the Treatment of Depression. & Journal of Clinical Psychiatry & 2018\\
    \hline
    3 & European Psychiatric Association Guidance on psychotherapy in chronic depression across Europe. & European Psychiatry & 2016\\
    \hline
    4 & Clinical practice guidelines on the evidence-based use of integrative therapies during and after breast cancer treatment. & CA: A Cancer Journal for Clinicians & 2017\\
    \hline
    5 & Depression: The Treatment and Management of Depression in Adults (Updated Edition). & National Collaborating Centre for Mental Health (UK) & 2010\\
    \hline
    6 & ACG Clinical Guideline: Preventive Care in Inflammatory Bowel Disease. & American Journal of Gastoenterology & 2017\\
    \hline
    7 & Screening for Depression in Children and Adolescents: US Preventive Services Task Force Recommendation Statement. & Pediatrics & 2016\\
    \hline
    8 & Management of Depression in Patients With Cancer: A Clinical Practice Guideline. & Journal of Oncology Practice & 2016\\
    \hline
    9 & Confronting depression and suicide in physicians: a consensus statement. & JAMA & 2003\\
    \hline
    10 & Screening for Depression in Adults: Recommendation Statement. & American Family Physician & 2016\\
    \hline
  \end{tabular}
\end{table*}

\begin{table*}
  \caption{Studies returned by CupQ for the query "stroke."}
  \label{tab:cupqstroke}
  \begin{tabular}{|l|p{10cm}|p{5cm}|l|}
    \hline
    \textbf{No} & \textbf{Title} & \textbf{Journal} & \textbf{Year}\\
    \hline
    1 & MRI-Guided Thrombolysis for Stroke with Unknown Time of Onset. & New England Journal of Medicine & 2018\\
    \hline
    2 & Clopidogrel and Aspirin in Acute Ischemic Stroke and High-Risk TIA. & New England Journal of Medicine & 2018\\
    \hline
    3 & Rivaroxaban for Stroke Prevention after Embolic Stroke of Undetermined Source. & New England Journal of Medicine & 2018\\
    \hline
    4 & Tenecteplase versus Alteplase before Thrombectomy for Ischemic Stroke. & New England Journal of Medicine & 2018\\
    \hline
    5 & Thrombectomy for Stroke at 6 to 16 Hours with Selection by Perfusion Imaging. & New England Journal of Medicine & 2018\\
    \hline
    6 & Thrombectomy 6 to 24 Hours after Stroke with a Mismatch between Deficit and Infarct. & New England Journal of Medicine & 2018\\
    \hline
    7 & Patent Foramen Ovale Closure or Antiplatelet Therapy for Cryptogenic Stroke. & New England Journal of Medicine & 2017\\
    \hline
    8 & Long-Term Outcomes of Patent Foramen Ovale Closure or Medical Therapy after Stroke. & New England Journal of Medicine & 2017\\
    \hline
    9 & Patent Foramen Ovale Closure or Anticoagulation vs. Antiplatelets after Stroke. & New England Journal of Medicine & 2017\\
    \hline
    10 & Cluster-Randomized, Crossover Trial of Head Positioning in Acute Stroke. & New England Journal of Medicine & 2017\\
    \hline
  \end{tabular}
\end{table*}

\begin{table*}
  \caption{Studies returned by PubMed for the query "stroke."}
  \label{tab:pubmedstroke}
  \begin{tabular}{|l|p{10cm}|p{5cm}|l|}
    \hline
    \textbf{No} & \textbf{Title} & \textbf{Journal} & \textbf{Year}\\
    \hline
    1 & Hereditary cerebral small vessel disease and stroke. & Clinical Neurology and Neurosurgery & 2017\\
    \hline
    2 & Imaging Markers of Post-Stroke Depression and Apathy: a Systematic Review and Meta-Analysis. & Neuropsychology Review & 2017\\
    \hline
    3 & Role of Total, Red, Processed, and White Meat Consumption in Stroke Incidence and Mortality: A Systematic Review and Meta-Analysis of Prospective Cohort Studies. & Journal of the American Heart Association & 2017\\
    \hline
    4 & Endarterectomy achieves lower stroke and death rates compared with stenting in patients with asymptomatic carotid stenosis. & Journal of Vascular Surgery & 2017\\
    \hline
    5 & The Course of Activities in Daily Living: Who Is at Risk for Decline after First Ever Stroke? & Cerebrovascular Diseases & 2017\\
    \hline
    6 & Prevalence, incidence, and factors associated with pre-stroke and post-stroke dementia: a systematic review and meta-analysis. & Lancet Neurology & 2009\\
    \hline
    7 & Acupuncture lowering blood pressure for secondary prevention of stroke: a study protocol for a multicenter randomized controlled trial. & Trials & 2017\\
    \hline
    8 & Decreased Serum Brain-Derived Neurotrophic Factor May Indicate the Development of Poststroke Depression in Patients with Acute Ischemic Stroke: A Meta-Analysis. & Journal of Stroke and Cerebrovascular Diseases & 2018\\
    \hline
    9 & Aerobic Exercises for Cognition Rehabilitation following Stroke: A Systematic Review. & Journal of Stroke and Cerebrovascular Diseases & 2016\\
    \hline
    10 & Types of stroke recurrence in patients with ischemic stroke: a substudy from the PRoFESS trial. & International Journal of Stroke & 2014\\
    \hline
  \end{tabular}
\end{table*}

\appendix

\end{document}